\documentclass[11pt]{article}
\usepackage{fleqn,cospar,epsfig}

\usepackage{fleqn,cospar}

\usepackage{graphicx}

\def\zycki{$\dot{\rm Z}$ycki}
\def\rozanska{R\^o$\dot{\rm z}$a\'nska}
\def\reffcosi{R$_{\rm eff} \sqrt{\cos i}$}
\def\ergs{ergs s$^{-1}$}
\def\sax{Beppo-SAX}
\def\rxte{RXTE}
\def\integral{INTEGRAL}

\def\comptt{{\sc CompTT}}
\def\rincost{{R$_{\rm in} \cos\theta$}}

\def\kte{kT$_{\rm e}$}

\def\eqw{EqW}

\def\mdot{$\dot{M}$}
\def\Rms{R$_{\rm ms}$}
\def\rin{R$_{\rm in}$}
\def\rg{R$_{\rm g}$}

\title{The broad band X-ray/hard X-ray spectra of accreting neutron stars}

\author{Didier Barret\address{Centre d'Etude Spatiale des
Rayonnements, 9 Avenue du Colonel Roche, 31028 Toulouse, FRANCE}}

\begin{document}

\maketitle
\begin{abstract}
  I review the energy spectra of low mass X-ray binaries (LMXBs)
  containing weakly magnetized accreting neutron stars (NS),
  emphasizing the most recent broad band (0.1-200 keV) spectral and
  timing observations performed by Beppo-SAX and RXTE. Drawing on the
  similarities between black hole candidate (BHC) and NS accretion, I
  discuss the accretion geometry and emission processes of NS LMXBs.
\end{abstract}

\section*{INTRODUCTION}
With Beppo-SAX and RXTE, for the first time since their discoveries
about 40 years ago, LMXBs with NS primary have been observed with good
sensitivities simultaneously from X-rays ($\sim 0.1$ keV) to hard
X-rays (up to $\sim 200$ keV).  These spectral observations have set
unprecedented constraints on the emission processes of these systems.
They have also shed some light on the accretion geometry in the
immediate vicinity of the NS, and revealed many new similarities
between BHC and NS accretion.  At the same time, the \rxte~fast timing
observations, in particular through the discovery and follow-up
studies of kilo-Hertz Quasi-Periodic Oscillations (kilo-Hz QPOs) have
provided direct diagnostics of the innermost regions of the accretion
disks (e.g.\ Van der Klis 2000).  Thus, in the last few years,
significant advances have been made on the study of the accretion
flows around NS. In the next section, I will briefly summarize the
state of the observations before the \sax~and \rxte~era.  Then, I will
review the most recent broad band spectral observations of NS. I will
then briefly present some results obtained from correlated spectral
and fast timing studies.  I will then discuss what we have recently
learned from all these results when combined together.

\section*{THE PICTURE BEFORE BEPPO-SAX AND RXTE ERA}
Prior to \sax~and \rxte, observations of LMXBs had been performed in
X-rays\footnote{X-rays/hard X-rays are defined as photons of energy
below/above $\sim 20-30$ keV.} with satellites such as EINSTEIN (e.g.
Christian \& Swank 1997), TENMA (Mitsuda et al.\  1984, 1989), EXOSAT
(e.g. White et al.\  1986, White et al.\  1988), GINGA (Mitsuda 1992),
more recently with ASCA (e.g. Narita et al.\  2001), and separately in
hard X-rays with SIGMA (e.g. Barret \& Vedrenne 1994, Churazov et al.\ 
1997), BATSE (e.g. Harmon et al.\  1996) and OSSE (e.g. Strickman et
al.\  1996).

The X-ray spectra of LMXBs were generally described as the sum of a
soft and hard component and were interpreted in the framework of two
models: the {\it eastern} model (also called the TENMA model, Mitsuda
et al.\  1984), and the {\it western} model (or EXOSAT model, White et
al.\  1988).  The soft component of the TENMA model is a multi-color
disk component approximating the emission from an optically thick
geometrically thin accretion disk.  The hard component is a weakly
Comptonized blackbody component.  Comptonization of seed photons
emitted at the neutron star surface/boundary layer would take place in
the inner parts of the accretion disks (Mitsuda et al.\  1989).  In the
EXOSAT model, the soft component is a single temperature blackbody,
attributed to an optically thick boundary layer, whereas the hard one
represents unsaturated Comptonization taking place in the innermost
regions of the accretion disks (White et al.\  1988).  In both models,
the luminosity of the component associated with the boundary layer was
systematically lower than the one attributed to the disk, 
contrary to theoretical expectations (Sunyaev \& Shakura 1986).

Luminosity related spectral changes were observed by TENMA from the
LMXB 4U1608-52 (Mitsuda et al.\  1989).  With decreasing luminosity,
the degree of Comptonization increased, and the X-ray spectrum
approached a power law shape.  In the low/hard states, deviations from
a simple power law, appearing as a broad absorption edge between 8 and
10 keV were first observed by GINGA from 4U1608-522 (Yoshida et al.\ 
1993).  These deviations could be interpreted as due to the reflection
of the incident power law by a relatively cold medium (Mitsuda 1992). 
Line emission centered between 6.4 and 6.8 keV, with equivalent width
of 70-130 eV and Full Width Half Maximum (FWHM) of $\sim 1$ keV had
also been reported (White et al.\  1986, see however Mitsuda 1992).


In hard X-rays, the observations suffered dramatically from the lack
of simultaneous X-ray coverage.  At the end of the SIGMA/CGRO era, we
knew that bright LMXBs (e.g. GX5-1) do not display significant hard
X-ray emission, although a variable and low luminosity hard X-ray tail
had been detected by OSSE from Sco X-1 (Strickman \& Barret 2000).  On
the other hand, a few low luminosity LMXBs had been detected up to 100
keV (Barret \& Vedrenne 1994, Churazov et al.\  1997).  Due to the
limited statistics of the data (the first detection was only at the
$5\sigma$ level!), the hard X-ray spectrum could be fitted by steep
power laws (index around 2.5-3.0), or thermal Bremmstrahlung with
temperatures around 50 keV, or by Comptonization models with electron
temperatures around 30 keV (e.g. Churazov et al.\  1997).  It was soon
hypothesized that the emission of a hard X-ray tail could be
associated with a low luminosity state in X-rays, when the X-ray
spectrum approaches a power law, and that the softness of the hard
X-ray spectrum could be due to the presence of an undetected break or
high energy cutoff (Barret \& Vedrenne 1994, Churazov et al.\  1997). 
This hypothesis was soon after confirmed by the detection by BATSE of
4U1608-522 quasi-simultaneously with GINGA during a low X-ray state
(Zhang et al.\  1996).  Thermal and non-thermal models had then been
proposed to account for the hard X-ray emission observed (for a review
see Tavani \& Barret 1997).

Over the last few years, significant advances have been possible
thanks to the broad band spectral capabilities of \sax~and \rxte.  Low
energy coverage (below $\sim 2$ keV) enables to resolve the soft
($\sim 1$ keV) component of the spectra, simultaneous hard X-ray
coverage is necessary to constrain the physical parameters of the hard
component.  In addition, a good overlap region between the X-ray and
hard X-ray bands is required to extract the Compton reflection
component.  Good spectral resolution below $\sim 10$ keV is then
needed to resolve the associated emission line and absorption edge
features.  The broad band spectral capabilities of \sax~are
illustrated in Frontera et al.\  (1998). \rxte~is presented in Bradt,
Swank \& Rothschild (1993).  In addition to providing broad band
coverage, \sax~and \rxte~are also very complementary satellites:
\sax~has good spectral resolution in X-rays, good sensitivity in hard
X-rays, and is perfectly suited for dedicated deep pointed
observations for detailed spectral studies.  On the other hand,
\rxte~with its large collecting area, high time resolution, and large
telemetry rate is {\it the} instrument of choice for fast timing
studies.  Furthermore, with its flexibility of operations, it offers
the unique capability of performing repeated observations over long
periods of time, allowing to sample multiple spectral and timing
states of the same source.  The results presented later should nicely
illustrate the unique capabilities of both satellites.

\section*{BROAD BAND BEPPO-SAX AND RXTE OBSERVATIONS}
\begin{figure}[!t]
\begin{minipage}{50mm}
  \hspace*{-1.0cm}\includegraphics[width=1.5\textwidth]{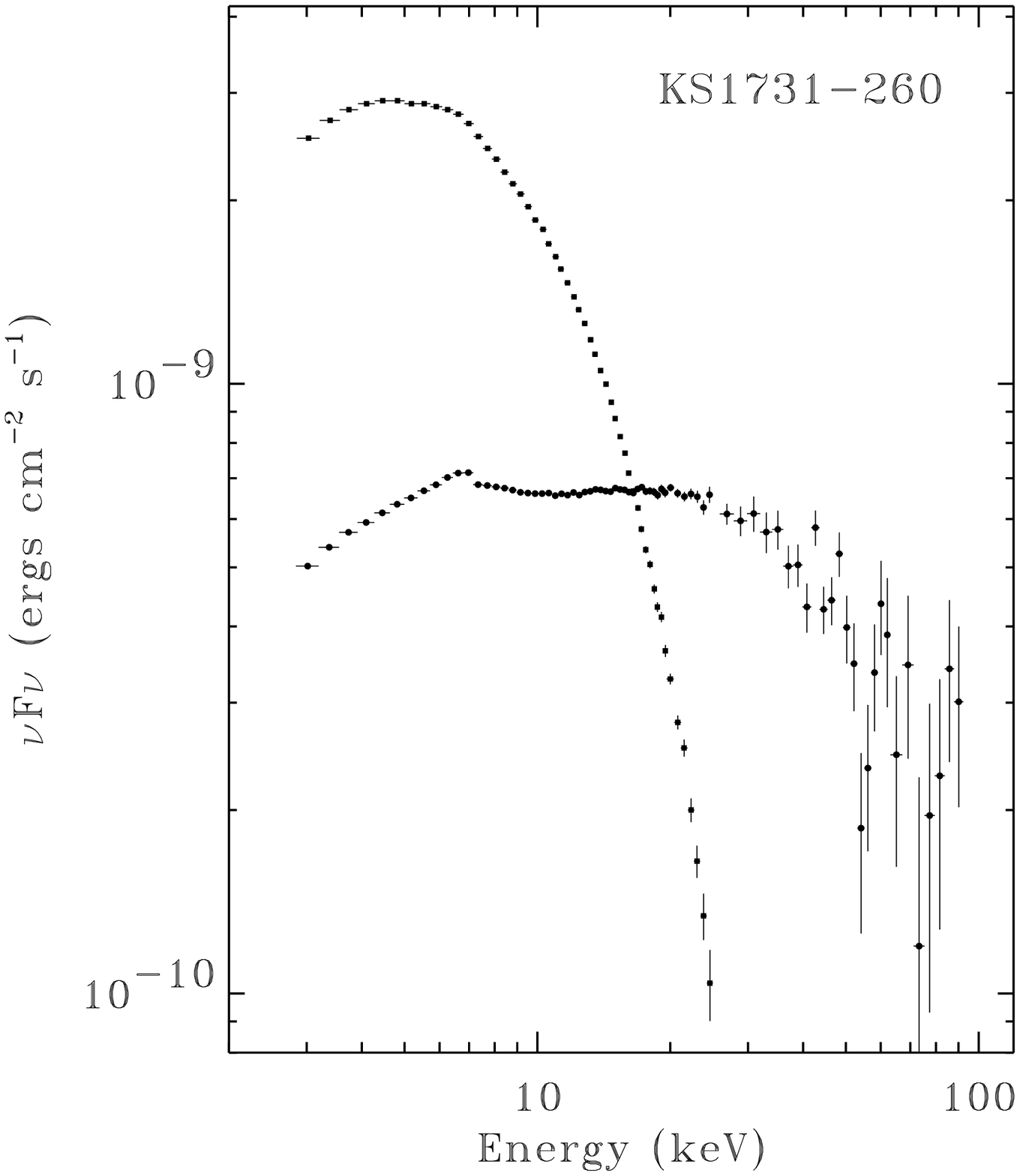}
\end{minipage} 
\begin{minipage}{50mm}
\hspace*{-0.15cm}\includegraphics[width=1.5\textwidth]{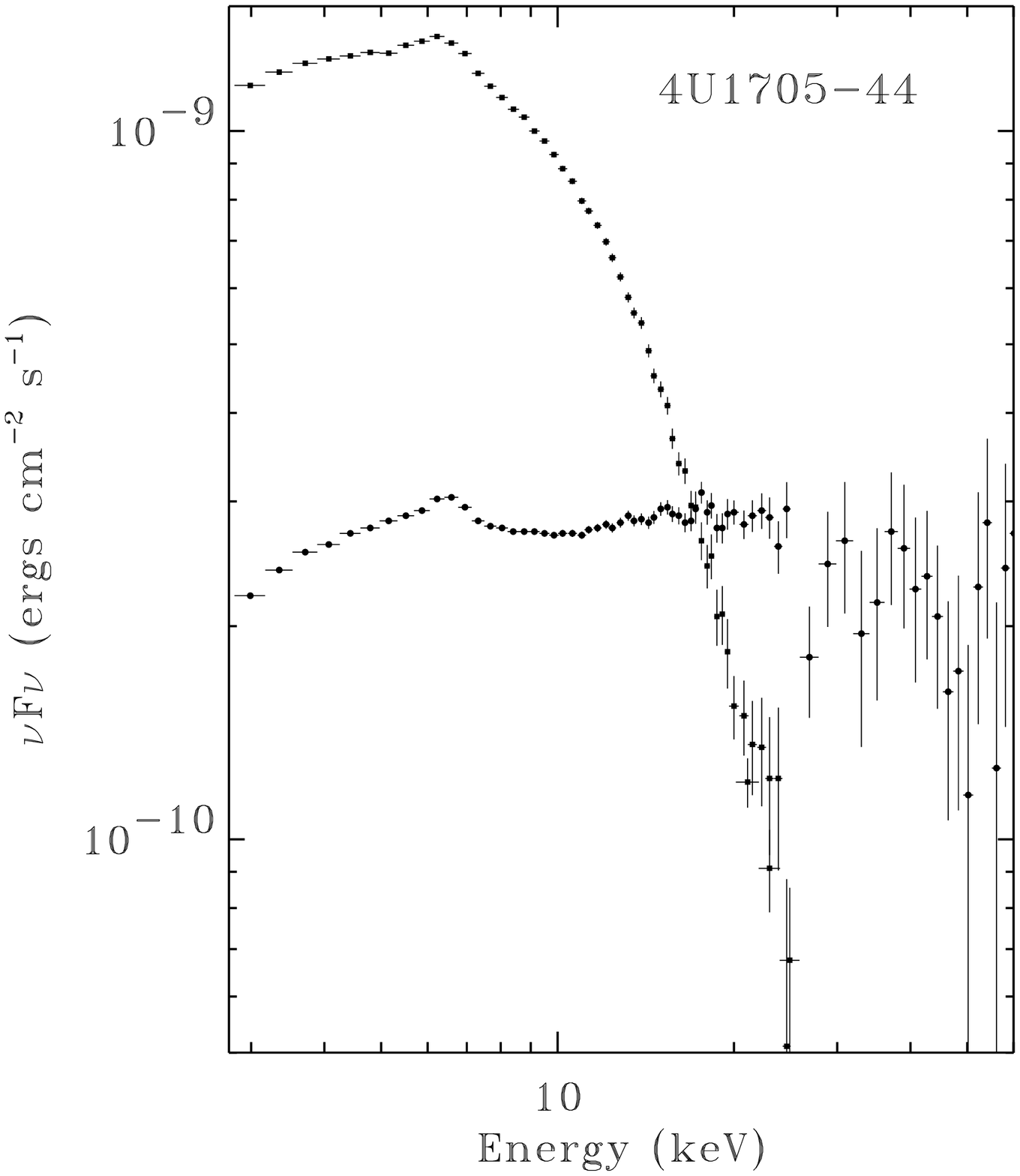}
\end{minipage} 
\begin{minipage}{50mm}
\hspace*{0.9cm}\includegraphics[width=1.5\textwidth]{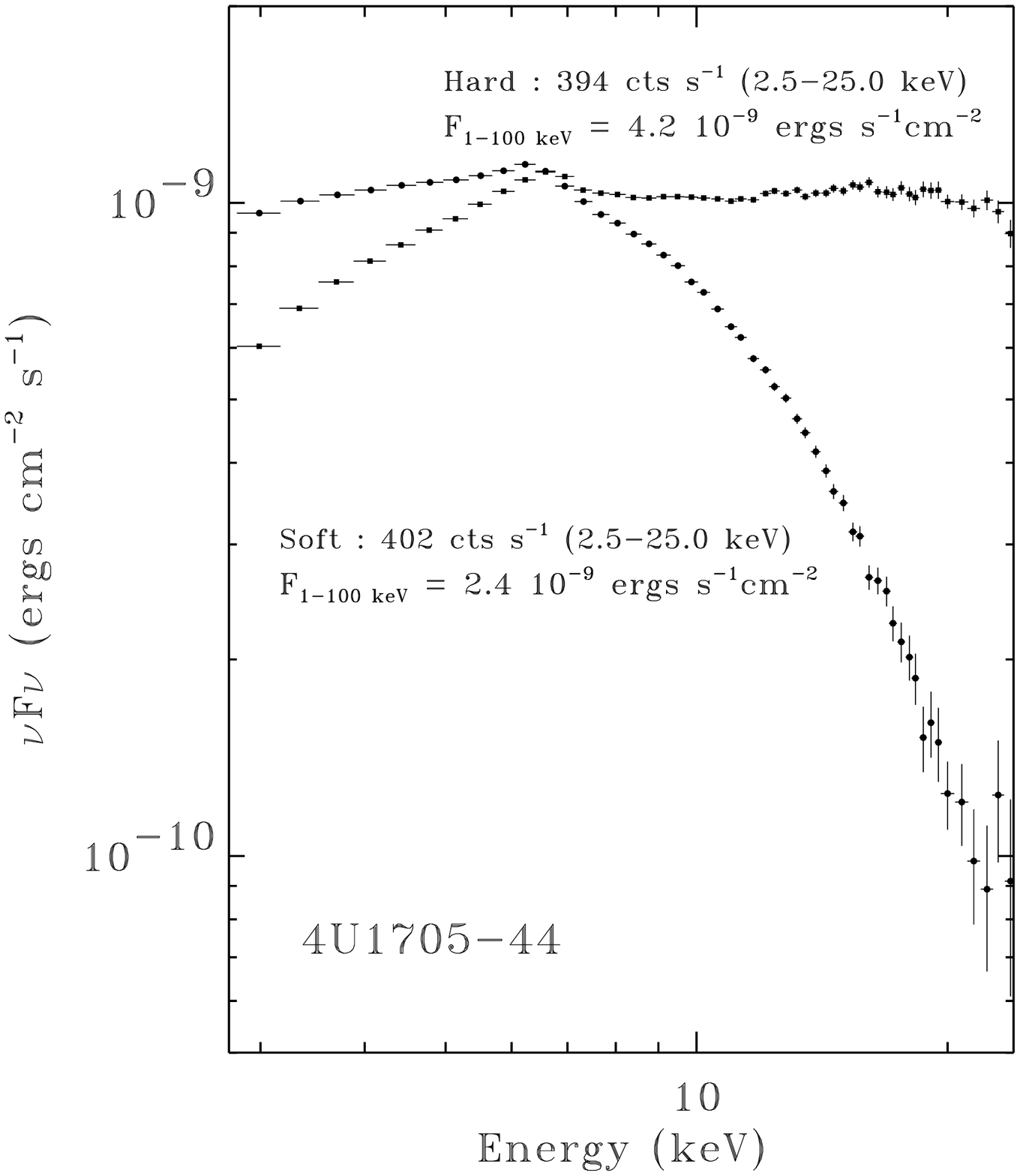}
\end{minipage} \vspace*{-1cm}
\caption{Spectral state changes observed by \rxte~from the two 
  variable LMXBs, KS1731-260 ({\it left}) and 4U1705-44 ({\it middle})
  as observed by \rxte. The soft spectra are well fitted by the sum of
  a MCD and a \comptt~whereas for the hard spectra they are well
  fitted by a BB/MCD and a cutoff power law (PCA and HEXTE spectra are
  combined). {\it Right)} Two PCA spectra taken by \rxte~from
  4U1705-44 during a spectral transition that occured in February
  1999.  The hard spectrum corresponds to a broad band luminosity that
  is about twice the one associated with the soft spectrum, owing to
  the appearance of a relatively strong hard X-ray component (the
  2.5-25.0 keV X-ray count rate differs by less than 2\% between the
  two observations, Barret \& Olive, work in preparation). This
  illustrates that the X-ray count rate alone is not a great indicator
  of the spectral state.}
\label{transitions} 
\end{figure}

Following on the detection of hard X-ray tails from low
luminosity\footnote{In the past, due to the absence of broad band
  coverage, the word luminosity usually referred to the X-ray
  luminosity.  In this paper, it refers to the broad band X-ray to
  hard X-ray luminosity.} LMXBs, Van Paradijs \& Van der Klis (1994)
showed, using the HEAO-1 A4 data that, in LMXBs there is a global
anti-correlation between the X-ray luminosity and spectral hardness.
Although the global trend is indeed real when considering the X-ray
luminosity (the higher X-ray luminosity sources have on average softer
spectra than lower luminosity sources), when considering broad band
luminosity the separation between soft and hard spectra is not as
clear.  There are some sources with soft spectra and low luminosities,
as well as some sources with hard spectra and relatively large
luminosities.  Furthermore, within a given source, it can be found
that a softer spectrum does not always imply a larger luminosity (see
Fig.  \ref{transitions}, right panel).  Finally the luminosity is
computed for a distance which is sometimes poorly constrained.  For
these reasons, in this paper, independently of the observed source
luminosity, I define the spectral hardness as a measure of the
fraction of flux radiated in the hard X-ray band.  A soft spectrum is
therefore a spectrum for which most (say larger than 80\%) of the
source flux is radiated below 20 keV.  Based on this, I will review
separately soft and hard spectra.  Both soft and hard spectral states
have been observed from a number of LMXBs characterized by large
intensity variations.  These spectral changes are sometimes very
spectacular as shown in Fig.  \ref{transitions} for the two LMXBs
KS1731-260 and 4U1705-44.

\subsection*{Soft spectra}
\begin{figure}[!t]
\begin{minipage}{60mm}
  \centerline{\includegraphics[width=.80\textwidth]{DBarret.fig2a.ps}}
\end{minipage} 
\begin{minipage}{60mm}
\centerline{\includegraphics[width=.83\textwidth]{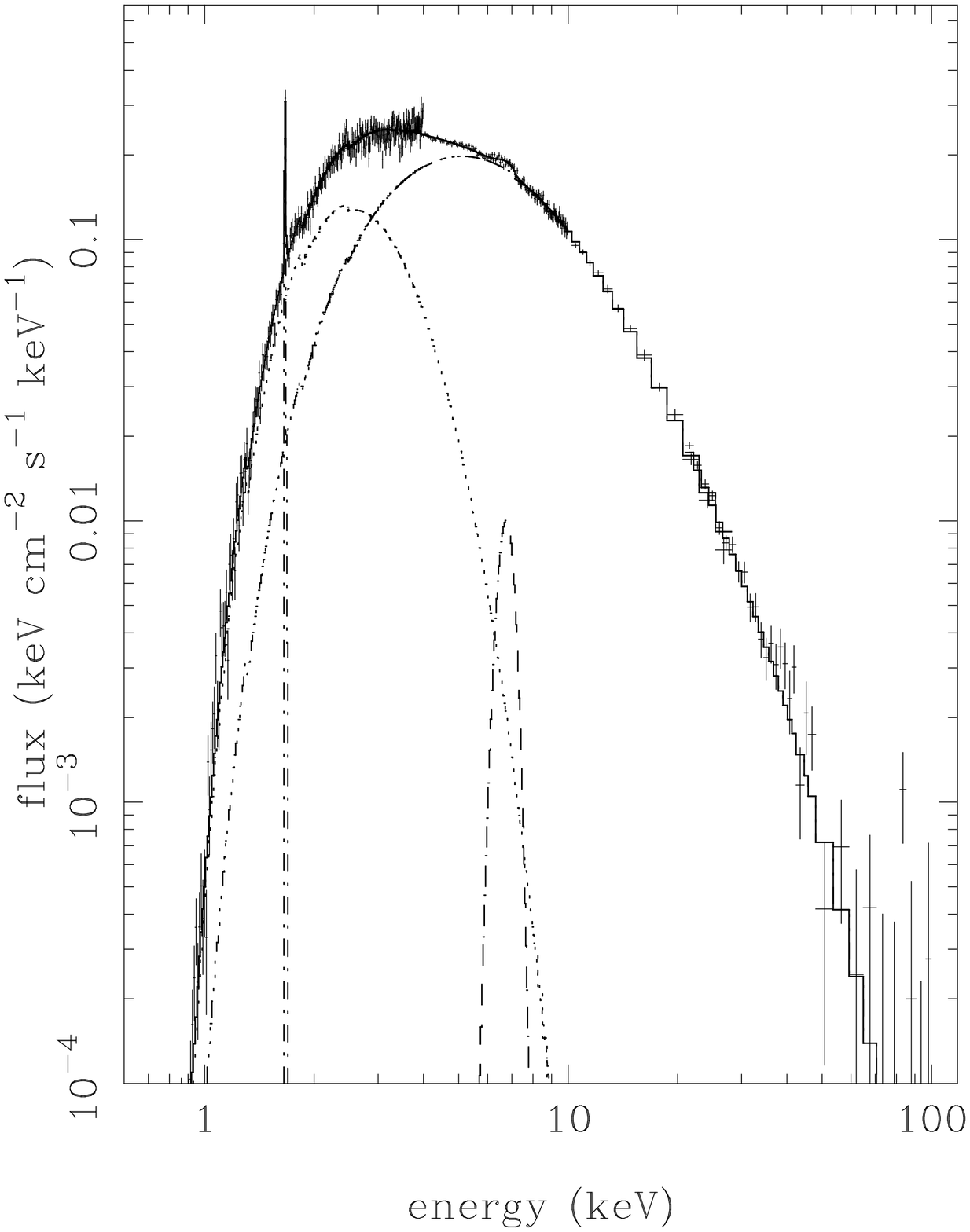}}
\end{minipage} 
\begin{minipage}{60mm}
\hspace*{-0.5cm}\centerline{\includegraphics[width=0.8\textwidth,angle=-90.]{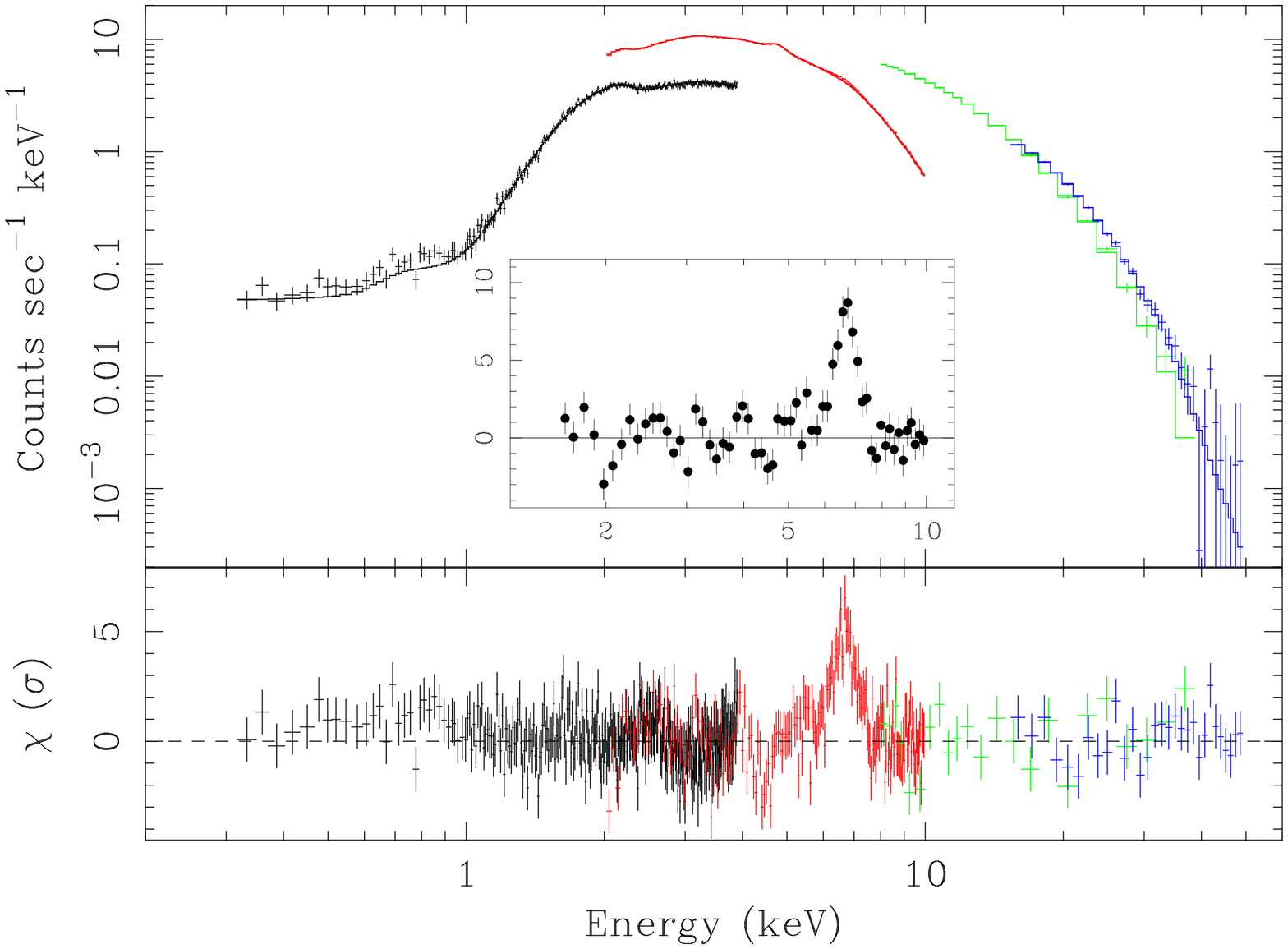}}
\end{minipage} 
\vspace*{-1.2cm} \caption{{\it Left)} A soft spectrum of Ser X-1 and
  {\it Middle)} 4U1728-34 as measured by \sax~(from Oosterbroek et al.\
  2001 and Di Salvo et al.\  2000a).  In both cases, the spectrum is the
  sum of a soft component (disk blackbody or blackbody) and a harder
  Comptonized component, plus a broad 6.4 keV and 6.7 keV line for Ser
  X-1 and 4U1728-34 respectively.  An additionnal line at 1.7 keV is
  detected in 4U1728-34. {\it Right)} The 0.1-50 keV soft spectrum of
  4U 1728-34 observed by BeppoSAX is shown together with the residuals
  in the entire band, in unit of standard deviations, when the best
  fit continuum is applied in the whole band except the 4-8 keV energy
  range. The inset shows the residuals of the MECS data rebinned to
  better display the profile of the observed Fe K$_\alpha$ feature
  (Piraino et al.\ 2000)}
    \label{sp_soft}
\end{figure}

Soft spectra have been observed over a range of luminosity going from
$\sim 10^{37}$ \ergs~to $\sim 10^{38}$ \ergs~(e.g. GX3+1, $1.3 \times
10^{37}$ \ergs~Oosterbroek et al.\ 2001, 4U1728-34, $1.8\times10^{37}$
\ergs~Piraino et al.\ 2000, GX17+2, $1.0-1.2 \times 10^{38}$\ergs~Di
Salvo et al.\ 2000). Soft spectra have been observed both from Z and
Atoll sources. These spectra are always decomposed as the sum of a
soft component and a Comptonized component (e.g.  4U1728-34, see Fig.
\ref{sp_soft}).  Despite its good sensitivity and spectral resolution
at low energies, which resolves nicely the soft component, Beppo-SAX
is not able to distinguish between a single temperature blackbody (BB)
and a multi-color disk (MCD) blackbody model (e.g. Di Salvo et al.\ 
2000b).  BB temperatures of less than 1 keV are generally observed.
For the MCD model the typical range for the color temperature is $\sim
0.5$ to 1.5 keV. In addition for the latter model, very small values
of the projected inner disk radius \rincost~are derived, typically a
few kilometers (e.g. 2.8 km for GX3+1, Oosterbroek et al.\ 2001).
Merloni et al.\ (2000) have shown however that the inner disk radius
so measured understimates the true inner disk radius (see also
Gierli{\'n}ski et al.\ 1999 for a discussion about the importance of
the assumed inner boundary conditions on the disk radius estimate).
The observed value is therefore poorly constraining, as it must be
corrected by a spectral hardening factor, which varies with the
accretion rate and the fraction of energy dissipated outside the disk
(Merloni et al.\ 2000).  With these limitations in mind, when
corrected for an invariant spectral hardening factor of 1.7, in the
best case, a plausible value of the effective inner disk radius is
obtained (e.g.  \reffcosi $\sim 20$ km in 4U1728-34, Di Salvo et al.\ 
2000a, $i$ is the inclination angle).

For the Comptonized component, a temperature of a few keV and a
relatively large optical depth of $\sim 5-15$ are observed~(e.g.
GX3+1, \kte=2.7 keV, $\tau=6.1$, Oosterbroek et al.\  2001, GX17+2,
\kte=3.0 keV,$\tau \sim 10$, Di Salvo et al.\  2000b, KS1731-260
\kte=2.7 keV, $\tau \sim 12$, Barret et al.\  2000).  Using the
\comptt~model in XSPEC (Titarchuk 1994), it is in principle possible
to determine the characteristic temperature of the seed photons for
the Comptonization.  Seed photon temperatures range from 0.3 to 1.5
keV, i.e. in the same range of temperatures measured for the soft
component (e.g. Oosterbroek et al.\  2001, Di Salvo et al.\  2000a)

The ratio between the fluxes of the soft (BB or MCD) and Comptonized
components varies typically between 0.1 and 0.5, thus indicating that
the soft component does not dominate the source luminosity. This has
generally led to the interpretation that the soft component originates
from an optically thick accretion disk, whereas the harder Comptonized
component arises from a hot inner flow and/or a hot boundary layer
with the seed photons coming from both the accretion disk and the NS
surface (e.g.  Di Salvo et al.\ 2000a, Barret et al.\ 2000).
Obviously, this is a revisited version of the previously mentioned
{\it eastern} model.

In addition to the above components, line features between 6.4 and 6.7
keV have been convincingly reported in a few NS soft spectra.  One of
the nicest example to date is provided by the \sax~observation of
4U1728-34 (Piraino et al.\  2000, see Fig \ref{sp_soft}, right).  The
line is broad (relativistic broadening ?)  and interpreted as emission
from highly ionized iron (Fe XXV to Fe XXXVI).  As an iron line
produced at the NS surface would be redshifted down to $\sim 5$ keV,
it most likely arise from an irradiated accretion disk (e.g.  Piraino
et al.\  2000).  In that case, one would expect also a reflection
component.  The reflection component is however not expected to be
intense, due to the softness of the primary spectrum, which means that
photons should be photo-electrically absorbed and thermalized in the
disk, rather than being Compton scattered.  Yet, weak evidence for the
presence of such a reflection component has been reported recently
(Oosterbroek et al.\  2000, Di Salvo et al.\  2000a).  In the case of
4U1728-34 (Fig. \ref{sp_soft}, Middle), using the reflection model of
\zycki~et al.\  (1998) in which the iron line is computed
self-consistently with the reflection component, the observed line
would require reflection from a moderately ionized disk ($\xi=280$, Di
Salvo et al.\  2000a).


\subsection*{Soft spectra and hard X-ray tails}
\begin{figure}[!t]
  \centerline{\includegraphics[width=0.4\textwidth]{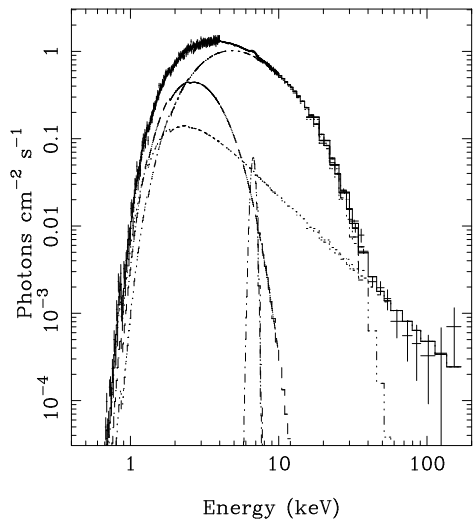}}
  \vspace*{-1.80cm}\caption{The unfolded spectrum of GX17+2 (Courtesy
    of T. Di Salvo).  The spectrum is fitted as the sum of a disk
    blackbody (MCD), an iron line, a Comptonized component and a steep
    hard X-ray tail of photon index 2.5 (Di Salvo et al.\ 2000b).}
    \label{gx17+2}
\end{figure}

Departing from the extrapolation of the X-ray spectrum around 40-60
keV, a hard X-ray tail extending out to 100 keV or more has been
detected with high significance in Cyg X-2 (Frontera et al.\ 1998), in
GX17+2 (Di Salvo et al.\ 2000b, Fig.  \ref{gx17+2}), in Sco X-1 by
HEXTE (D'Amico et al.\ 2000) thus confirming the previous detections
by OSSE (Strickman \& Barret 2000), in GX349+2 (Di Salvo et al.\ 
2001a), and from the somewhat peculiar Cir X-1 source (Iaria et al.\ 
2000).  In Sco X-1, the luminosity of the hard tail is less than 1\%
of the broad band source luminosity, whereas it reaches 8\% in GX17+2.
Due to the limited statistics, these hard tails have been fitted by
simple power laws.  The photon index measured by \sax~is 2.5 and 2.0
for GX17+2 and GX349 respectively (Di Salvo et al.\ 2000b, 2001a).  Di
Salvo et al.\ (2000b) have shown that for GX17+2 the hard tail was
detected only at the lowest {\it inferred} mass accretion rate (i.e.
when it is on the so-called horizontal branch of the color-color
diagram, CCD).  This contrasts to the case of Sco X-1, for which
D'Amico et al.\ (2000) have shown that the presence of a hard tail was
not confined to a specific position of the source in its CCD. The lack
of clear energy cutoffs in the observed hard tails (this must be
confirmed by higher sensitivity observations, e.g. with INTEGRAL),
together with the fact that the sources above are relatively bright
radio sources (Fender \& Hendry 2000) has led to the hypothesis that
the hard X-ray emission could have a non thermal origin (see
discussion below).

The spectrum shown in Fig \ref{gx17+2} shares clearly some
similarities with those of high state BHCs, when the broad band
spectrum is the sum of a so-called {\it ultrasoft} component and a
hard X-ray tail (Grove et al.\ 1998).  There are two noticeable
differences however.  First in NS, as we have shown above the X-ray
spectrum is the sum of MCD/BB and a harder Comptonized component,
whereas for BHC the X-ray spectrum can be generally modeled by a
single component, approximated by a MCD (e.g. Tanaka \& Lewin 1995).
This difference makes the X-ray part of the spectrum of NS to look
{\it harder} than the one for BHCs.  Second, concerning the hard X-ray
tail, in BHCs it is most of (if not all) the time present, whereas the
available data suggest that it is much more variable in NS.


\subsection*{Hard spectra}
Hard spectra have been observed for luminosities ranging from a few
times $10^{36}$ \ergs~up to $\sim 3 \times 10^{37}$ \ergs.  So far
only Atoll sources (and more generally X-ray bursters) have been
observed with hard spectra. For those spectra, about half of the
source luminosity is radiated in hard X-rays.  There are now about 20
NS LMXBs detected up to 100 keV (Barret et al.\ 2000).  For all of
them and within distance uncertainties, the hard X-ray luminosity
never exceeds $\sim 1.5 \times 10^{37}$ \ergs.  This is unlike for
BHCs for which the hard X-ray luminosity can largely exceed that value
(Barret et al.\ 2000).

\begin{figure}[!t]
\begin{minipage}{80mm}
\hspace*{0.75cm}\centerline{\includegraphics[width=.75\textwidth,angle=-90]{DBarret.fig4a.ps}}
\end{minipage}
\begin{minipage}{80mm}
\hspace*{1.5cm}\centerline{\includegraphics[width=0.9\textwidth]{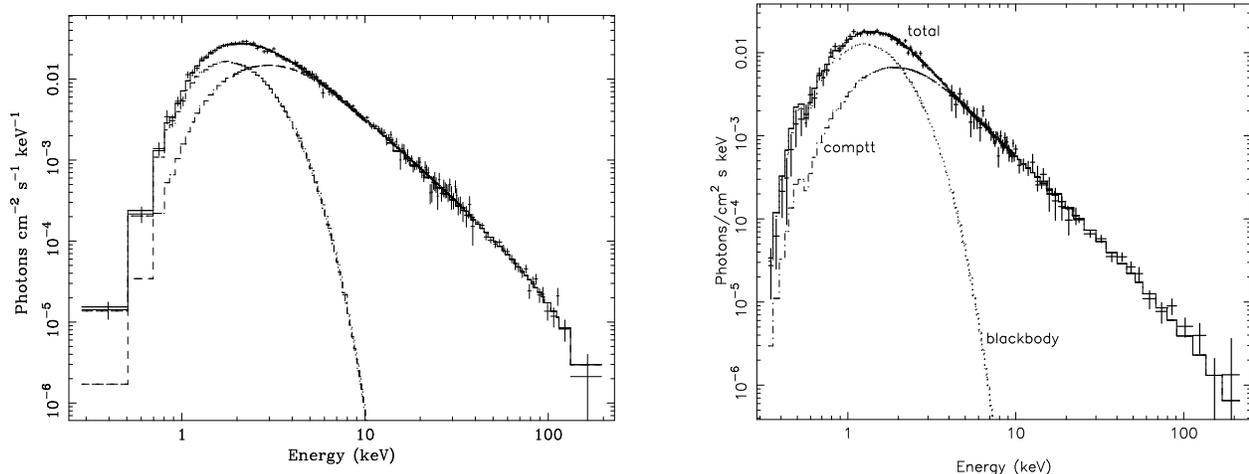}}
\end{minipage}
\vspace*{-1cm} \caption{{\it left)} The unfolded spectra of
  1E1724-3045 in Terzan 2 (Guainazzi et al.\  1998) and {\it right)}
  SAXJ1810.8-2609 (Natalucci et al.\  2000a).  In both cases, the hard
  spectrum can be fitted by the sum of a soft component (either a BB
  or a MCD) and a hard Comptonized component (for SAXJ1810.8-2609
  however, the hard X-ray component can also be fitted with a simple
  power law).}
\label{hardspectra}
\end{figure}

Like for soft spectra, broad band hard spectra are generally described
by the sum of a soft component and a Comptonized component.  The main
difference between the two classes of spectra resides in the
parameters of the Comptonizing component.  The inferred optical depth
of the Comptonizing cloud is now a few ($\sim 2-3$ for a spherical
geometry), whereas the electron temperature is typically a few tens of
keV, equivalent to an energy cutoff around 60-80 keV in the spectra
(e.g.  1E1724-3045, \kte=27 keV, $\tau$=3.3, Guainazzi et al.\ 1998,
SAXJ1747.0-2853, \kte=33 keV, $\tau$=3.1, Natalucci et al.\ 2000a, see
Fig.  \ref{hardspectra}).  The above values are typical of NS LMXBs,
and seem to be lower than the values observed from BHCs (e.g.
\kte~100 keV for Cyg X-1, Di Salvo et al.\ 2001b, see also Natalucci
2001).  The idea the electron temperature could be used as a criteria
for distinguishing between BHC and NS has already been invoked (Tavani
\& Barret 1997, Zdziarski et al.\ 1998, Barret et al.\ 2000), and
interpreted as the signature of the neutron star surface acting as a
thermostat for the Comptonization region (e.g. Klu\'zniak 1993).

\begin{figure}[!t]
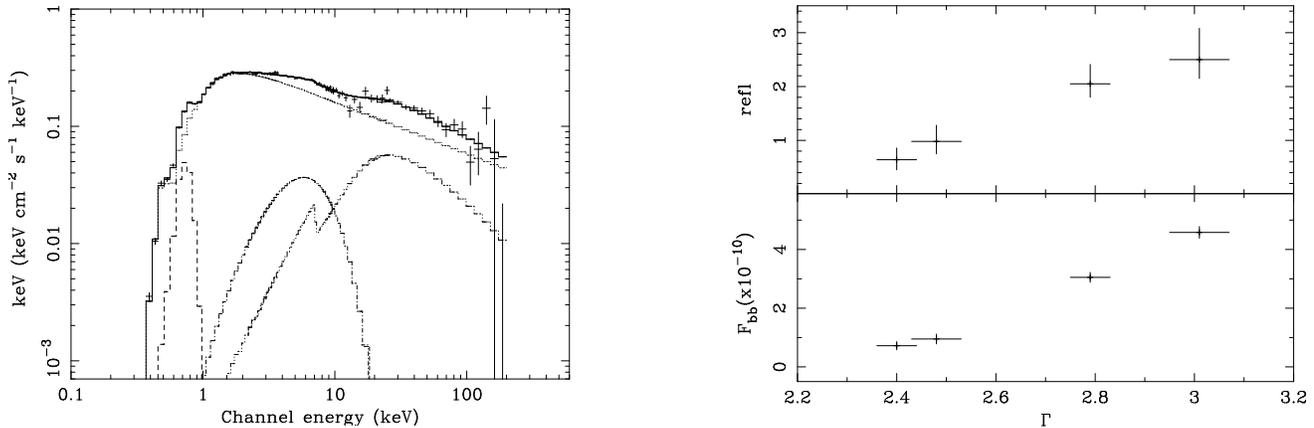

\begin{minipage}{80mm}
\includegraphics[width=.70\textwidth,angle=-90]{DBarret.fig5a.ps}
\end{minipage}
\begin{minipage}{80mm}
\hspace*{1.5cm}\includegraphics[width=.70\textwidth,angle=-90]{DBarret.fig5b.ps}
\end{minipage}
\vspace*{-0.8cm} \caption{{\it left)} \sax~unfolded averaged spectra
  of 4U0614+09, for the October 19--20, 1998 observation together with
  a model consisting of a powerlaw, reflection, a blackbody, and a low
  energy Gaussian line all with absorption.  The total model fit is
  shown as a solid line, the powerlaw as dotted line, the reflection
  component as a dot-dot-dashed line, the blackbody component as a
  dot-dashed line, and the Gaussian line as a dashed line.  No clear
  high energy cutoff is observed in the hard X-ray tail.  {\it Right)}
  Spectral parameters correlations.  (a) Magnitude of reflection (as
  defined above) versus photon index.  (b) Blackbody flux ($\rm erg\;
  cm^{-2}\; s^{-1}$) versus photon index (From Piraino et al.\ 2000).}
\label{reflection}
\end{figure}

However, in a few cases so far, no clear energy cutoffs were observed
in the hard X-ray spectrum (e.g. Aql X-1, Harmon et al.\  1996).  For
4U 0614+09 a lower limit of 220 keV on \kte~was derived (Piraino et
al.\  1999).  These non attenuated power laws have photon index in the
range 2.0-2.5 (see Fig.  \ref{reflection}, left).  They clearly share
some similarities with those observed during the so-called power-law
gamma-ray spectral state of BHCs as defined in Grove et al.\  (1998). 
The lack of high energy cutoffs may again be a signature of
non-thermal Comptonization (Poutanen 1999 for a review).

As far as the soft component is concerned, it can be fitted either by
a blackbody or a multi-color blackbody and contributes modestly to the
source luminosity (e.g. 10\% in SAXJ1747.0-2853, Natalucci et al.\ 
2000b, 15-30\% in 1E1724-3045, Barret et al.\ 2000).  Sometimes the
soft component is not detected (e.g. In't Zand et al.  1999 for
SAZJ1748.9-2021), this may be because it is absent, or too faint, or
that the bulk of its emission is radiated below the observed energy
range.

Below the continuum, a reflection component has been detected in a few
cases (e.g.  SAXJ1808-3658; Gierli{\'n}ski et al.\  2001, GS1826-238,
Barret et al.\  2000, see Fig.  \ref{reflection}).  This reflection
component is important as it can be used to probe the ionization
state, element abundances and geometry of the accretion flow (e.g.
Georges \& Fabian 1991, Matt et al.\ 1991).  In general it is found
that the reflector is neutral/moderately ionized (e.g. Barret et al.\
2000). Furthermore the magnitude of the reflected component
(R\footnote{Compton reflection is generally measured as a relative
  normalization ($R$) of the reflected component with respect to the
  irradiating component.  If the irradiating source is isotropic and
  neither the primary nor the reflecting medium are obscured, then
  R=$\Omega/2\pi$, where $\Omega$ is the solid angle subtended by the
  reflector as viewed by the irradiating source.}) indicates that the
reflecting medium subtends a small solid angle to the irradiating
source.  Consistent with the presence of reflection, a weak iron line
has been detected (\eqw$\sim 50$ eV).  These observed values are
significantly lower than the $\sim 130$ eV and $R=1$ of an isotropic
X-ray source above a flat infinite slab (Georges \& Fabian 1991).  The
most likely site for the reflection is the accretion disk.  However,
for 4U0614+09 for which no thermal cutoffs were observed in the hard
X-ray tail, a stronger reflection component has been observed with R
ranging from 1 to $\sim 3$.  Such a large $R$ implies that the primary
source is either partially obscured or anisotropic in the reflector
frame. This might be expected if the power law results from
non-thermal Comptonization on relativistic electrons (Piraino et al.\
1999).  In addition, a correlation between the strength of the
reflection component and the index of the power law has been found
(see Fig.  \ref{reflection}).  A similar correlation has been found in
Seyferts AGNs, BHC and the X-ray bursters (GS1826-238 and 4U1608-522),
thus suggesting that a similar accretion geometry occur in a wide
range of systems, independently of the nature and mass of the compact
object (Zdziarski et al.\  1999).  This correlation can be interpreted
if the reflecting medium (most likely the accretion disk) plays a
dominant role as a source of seed photons for the Comptonization in
the irradiating source (Zdziarski et al.\  1999).


\begin{figure}[!b]
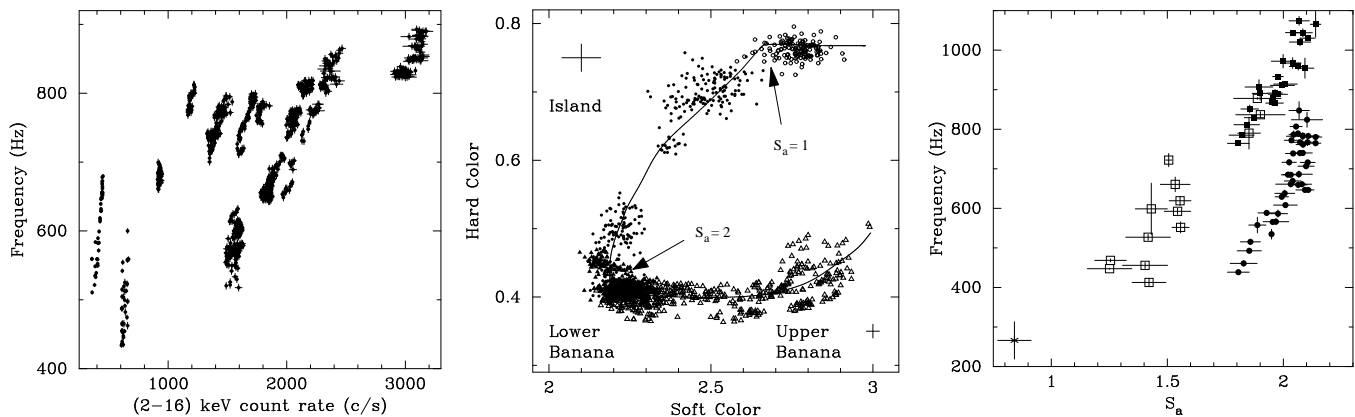

\begin{minipage}{60mm}
\centerline{\includegraphics[width=.9\textwidth,angle=-90]{DBarret.fig6a.ps}}
\end{minipage}
\begin{minipage}{60mm}
\centerline{\includegraphics[width=.9\textwidth,angle=-90]{DBarret.fig6b.ps}}
\end{minipage}
\begin{minipage}{60mm}
\centerline{\includegraphics[width=.9\textwidth,angle=-90]{DBarret.fig6c.ps}}
\end{minipage}
\vspace*{-0.8cm}
\caption{{\it Left)}: Relation between $\nu_{1}$ and the $2-16$
  keV count rate in 4U\,1608--52 ($\nu_{1}$ is associated here with
  the lower kilo-Hz QPO).  {\it Center)} X-ray color-color diagram.
  The soft and hard colors are defined as the ratio of count rates in
  the bands $3.5 - 6.4$ and $2.0 - 3.5$ keV, and $9.7 - 16.0$ and $6.4
  - 9.7$ keV, respectively. The curve shows the parametrization of the
  color-color diagram in terms of S$_{\rm a}$ The accretion rate onto
  the NS is supposed to increase along S$_{\rm a}$ {\it right)}: For
  the same data as on the left panel, $\nu_{1}$ (filled circles) and
  $\nu_{2}$ (squares both opened and filled) versus the previously
  defined S$_{\rm a}$.  Adapted from M\'endez et al.\ (1999).}
\label{mendez99}
\end{figure}

\section*{CORRELATED TIMING AND SPECTRAL STUDIES}
Kilo-Hz QPOs have now been detected from over 20 LMXBs (Van der
Klis 2000).  Without entering into the details of the QPO
phenomenology, in most sources two kilo-Hz QPOs have been observed.
Although the origin of the lower QPO is still debated, in most models,
the higher QPO is associated with a Keplerian frequency at the inner
edge of the accretion disk (Van der Klis 2000).  Thus for a NS
producing kilo-Hertz QPO, some new information can be gathered by
relating the luminosity/spectral changes observed to changes of the
QPO frequency.

On short time scales (a few hours to less than 1 day) there is a good
correlation between the kilo-Hz QPO frequency and the X-ray count
rate.  On longer time scales, the correlation does not hold, and while
the source span the same frequency range, its count rate can change by
up to a factor of a few (e.g M\'endez et al.\ 1999, see Fig.
\ref{mendez99}, left).  On the other hand, a much better correlation
on all time scales exists between the kilo-Hz QPO frequency and the
X-ray spectral shape, when measured as a position in the X-ray
color-color diagram (M\'endez et al.\ 1999, Fig.  \ref{mendez99},
right).  Kilo-Hz QPOs correlate well with some other timing features
(e.g. break frequency in the power density spectrum; e.g. Ford et al.\ 
1997), all better correlating with the X-ray colors than with the
X-ray flux.  From X-ray burst studies, simultaneous optical/UV and
X-ray observations, the sense of variation of \mdot~on the color-color
diagram has been determined. For instance in Atoll source the
\mdot~should increase from the so-called {\it Island} state to the
{\it upper-banana} state (see Fig. \ref{mendez99}, e.g. Van der Klis
1994).  This has led to the definition of an \mdot~inferred from the
position of the source on the color-color diagram.  Kilo-Hz QPOs, like
other timing features appear therefore to be set by this inferred
\mdot.  If this inferred \mdot~is the total mass accretion rate, then
the lack of correlation between the X-ray flux and inferred
\mdot~needs to be explained.  Several arguments have been put forward;
e.g. a time variable beaming of the emission, or a redistribution of
energy over an unobserved energy range, or the existence of
significant time variable mass outflows to remove mass and kinetic
energy (see e.g. Ford et al.\ 2000).  Alternatively, following an idea
developed in Fortner et al.\ (1989) (see also Lamb 1989), it has been
suggested that the accretion flow could be decoupled in the form of a
disk and radial\footnote{The mechanism to produce such a flow is the
  following: For sufficiently high luminosities, the radiation
  pressure from the disk drives some gas into a corona above the
  disk. Radiation drag causes the gas to lose its angular and
  verticular momentum and to fall radially toward the star (Fortner et
  al.\ 1989, Lamb 1989)} flow, and that the inferred \mdot~could be
the accretion rate through the disk (e.g.  Kaaret 2000, see discussion
below). In that picture, the total (disk plus radial) mass accretion
rate would determine the X-ray flux, which could in turn be decoupled
from any other observed quantities, such as the frequency of the
kilo-Hz QPOs.
\begin{figure}[!b]
\begin{minipage}{60mm}
\centerline{\includegraphics[width=.90\textwidth]{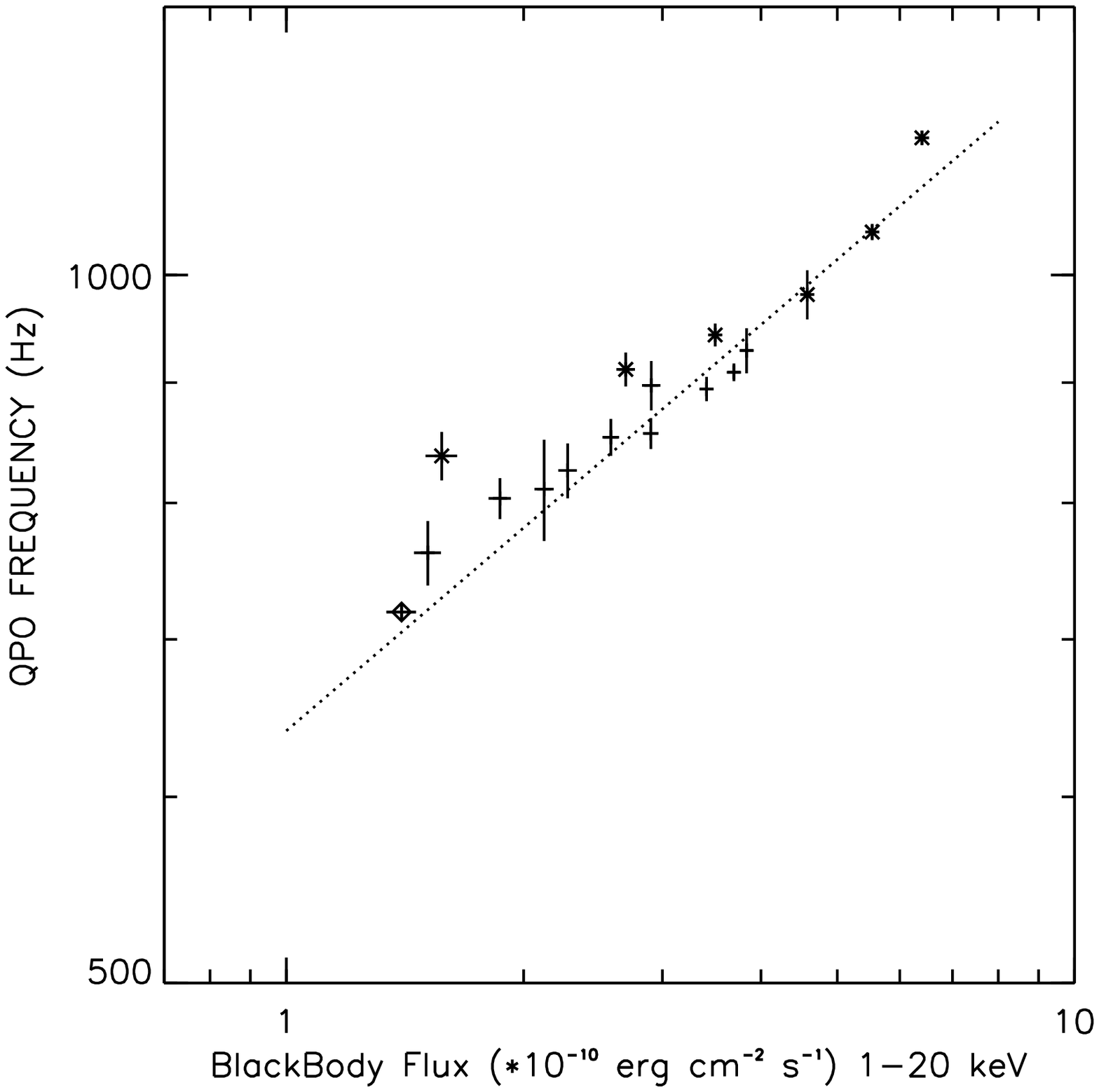}}
\end{minipage}
\begin{minipage}{60mm}
\centerline{\includegraphics[width=.90\textwidth]{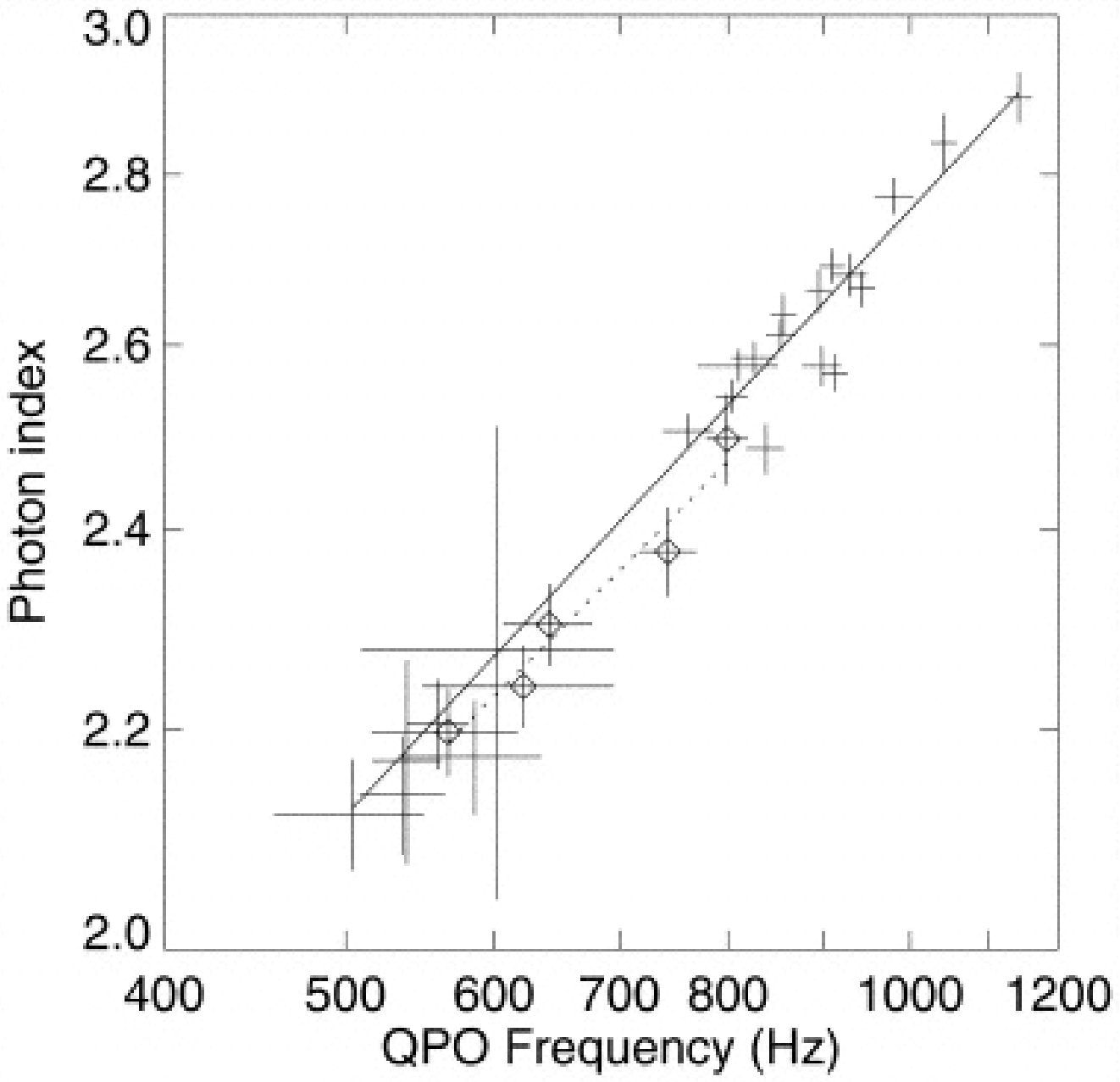}}
\end{minipage}
\begin{minipage}{60mm}
\centerline{\includegraphics[width=1.1\textwidth]{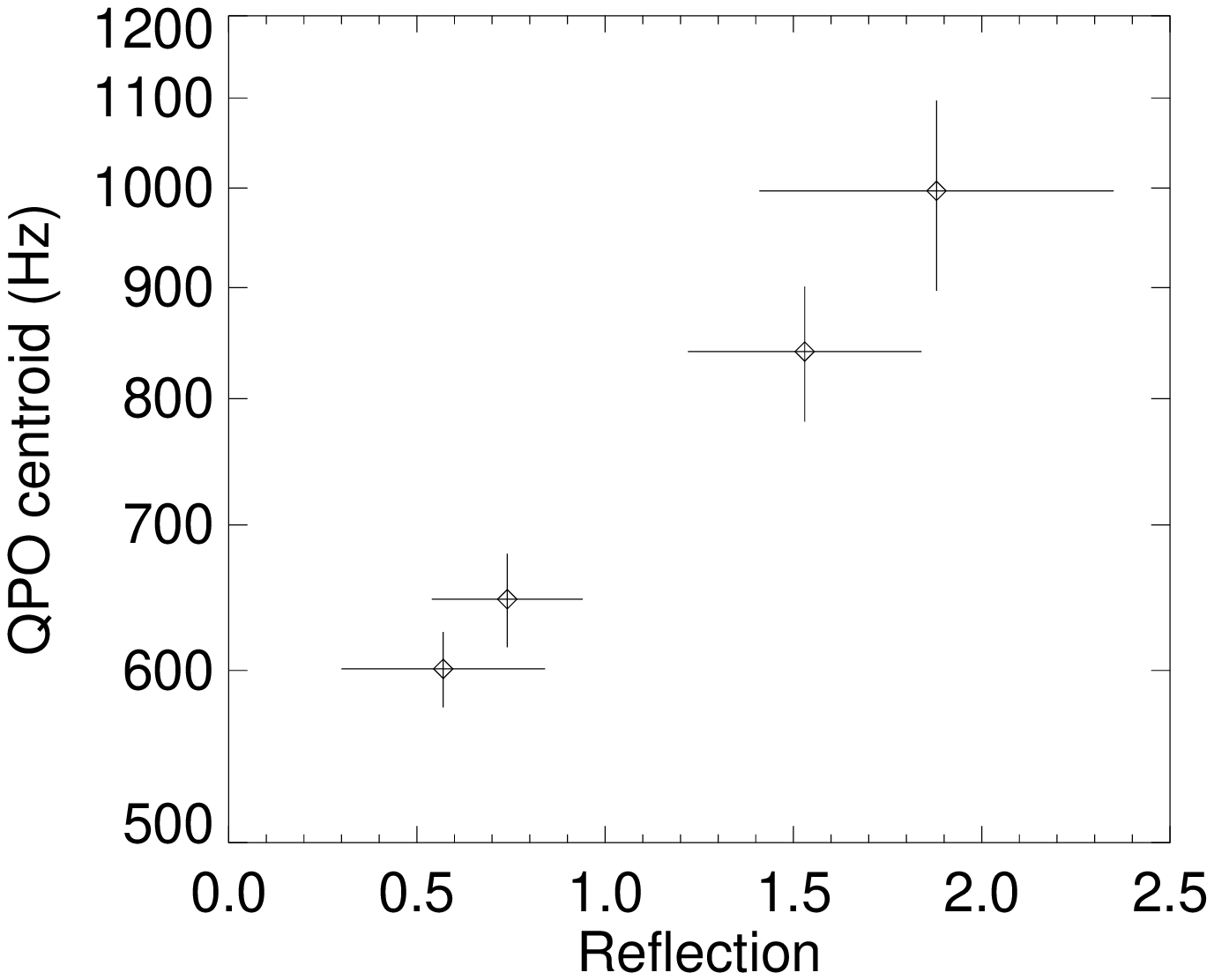}}
\end{minipage}
\vspace*{-1cm}
\caption{{\it Left)} Kilo-Hz QPO frequency versus  flux of the blackbody
  spectral component (Ford et al.\ 1997). {\it Middle)} Photon index vs.
  QPO frequency on a log-log plot.  The crosses indicate data for 4U
  0614+09; the solid line is a power-law fit to the data for 4U
  0614+09; the diamonds and the dashed line indicate the data and
  fit, respectively, for 4U 1608-52 (Kaaret et al.\ 1998). {\it Right)}
  Correlation between the strength of the reflection component and the
  QPO frequency in 4U0614+09 (Kaaret 2000)}
\label{correlations}
\end{figure}

When considering the components of the X-ray spectrum rather than
X-ray colors, the correlation between kilo-Hz QPOs and spectral shape
can be made very clear.  Ford et al.\ (1998) have shown that there is
a one to one correlation between the flux of the blackbody component
of the X-ray spectrum and the QPO frequency (Fig. \ref{correlations}
left). No such correlation exists between the total X-ray flux and the
QPO frequency.  Similarly, when fitting the X-ray spectra of 4U0614+09
and 4U1608-52 with simple power laws, Kaaret et al.\ (1998) observed a
correlation between the index of the power law and the QPO frequency,
the larger the index, the higher the QPO frequency (Fig.
\ref{correlations} middle).  Finally, Kaaret (2000) have shown that
the strength of the reflection component (Fig. \ref{reflection}) also
correlates with the QPO frequency (Fig.  \ref{correlations} right).
\section*{DISCUSSION}

The list of spectral similarities between BHC and NS accretion has
grown considerably over the last few years.  Some new similarities are
illustrated in this paper: non-thermal power law tail, Compton
reflection, soft X-ray spectrum and steep hard X-ray tail.  These
spectral similarities add to the long list of already known
similarities of the timing properties of the two classes of accreting
sources (Van der Klis 1994, see Wijnands 2001 for a recent review of
timing properties of LMXBs).  The similarities are especially striking
in the hard spectral state of those sources, when they both display
large amplitude band limited noise (flickering), and hard energy
spectra.  This implies that the timing and spectral properties of NS
are weakly affected by the presence of a hard surface (and a boundary
layer) or a small magnetosphere and that very similar accretion flows
exist around BHC and NS.

\subsection*{Accretion geometry and emission processes}
For BHC, the available spectral and timing data are consistent with
two accretion geometries (see Done 2001 for an extensive discussion).
The first one consists of a truncated optically thick geometrically
thin accretion disk, beyond which the nature of the flow changes to
form a hot inner disk region, most probably an advection dominated
accretion flow (ADAF, Narayan 1997 for a review).  The broad band
spectrum of Cyg X-1, the low amplitude of reflection and relativistic
smearing would be consistent with a truncation radius of the order of
a few tens of \rg~(\rg=$2GM/c^2$) (Di Salvo et al.\ 2001b).
Similarly, frequency resolved spectroscopy of Cyg X-1 yields an inner
disk radius of $\sim 100$ \rg~in the hard state and less than $\sim
10$ \rg~in the soft spectral (high) state (Gilfanov et al.\ 2000).  In
that picture, the soft component of the spectrum comes from the
truncated accretion disk, whereas the hard component arises through
thermal Comptonization from the hot inner disk region.  Conduction of
heat between the hot inner flow and a cold disk leading to the
evaporation of the disk has been proposed as a mechanism for the
transition (\rozanska~ \& Czerny, 2000).

The alternative geometry is an accretion disk extending all the way to
the last stable orbit, with the hard X-rays produced in active regions
above the disk, most likely powered by magnetic reconnection (e.g.
Haardt et al.\  1994).  To be reconciled with the spectral observations
(mainly the weakness of the reflection and reprocessed component), the
active regions should be expanding away from the disk with
relativistic velocities (Beloborodov 1999), or the disk should be
strongly photo-ionized (Ross et al.\  1999, Nayakshin et al.\ 2000).
Magnetic flares above a disk could account for the fast variability
and the magnitude of the hard time lags observed in both BH and NS
(e.g.  Ford et al.\ 1999), while keeping the X-ray source small
(Poutanen \& Fabian 2000).

In NS, if one assumes that the disk terminates where the kilo-Hz QPO
is produced, then the disk is truncated.  Kilo-Hz QPO range typically
from $\sim 300$ to 1300 Hz which correspond to Keplerian orbital
radius of 15 to 50 km.  From both observational (Psaltis et al.\ 1999)
and theoretical grounds (e.g. Stella et al.\ 2000), there is some
evidence that those QPOs could be produced at even lower frequencies
($\sim 100$ Hz or less) for lower luminosity sources, in which case,
this would push the inner disk radius much further out ($\sim 100$
km).  So if one assume that the kilo-Hz QPOs are produced at the
truncation radius between the cool disk and the hot inner disk region
then the correlations described above (e.g. Fig.  \ref{correlations})
can be naturally explained.  This is because \rin~not only sets the
QPO frequency, but determines also the spectral shape, as the
accretion disk is the main source of cool photons for the
Comptonization (Kaaret 2000).  For instance, when \rin~decreases, the
kilo-Hz QPO frequency increases, and the disk blackbody flux increases
(Ford et al.\ 1997).  At the same time, the spectrum steepens in
response to an increased cooling flux from the disk (see Fig.
\ref{correlations}, note that this will be true only if the spectrum
in produced through thermal Comptonization). Simultaneously, the angle
subtended by the truncated disk to the hot inner disk region
increases, leading to an increase of the amplitude of the reflection
component\footnote{The profile of the Fe line associated with the
  reflection component could also be used to test the proposed
  picture. When the disk moves in, the combination of Doppler effects
  and gravity in the vicinity of the NS should lead to appreciable
  changes in the line profiles. Similarly, as shown by Nayakshin
  (2000) in the accretion disk model with an X-ray heated skin, the
  parameters of the Fe line and absorption features should also change
  when the spectrum changes; harder spectra are expected to produce no
  Fe lines and no edges, whereas softer spectra should yield stronger
  ionized absorption edges and highly ionized Fe lines. This clearly
  emphasizes the need for spectroscopic observations (e.g. with
  XMM-Newton and Chandra) of those systems.} (Kaaret 2000).  It has
then been argued that, provided that the accretion takes place through
two different channels (as discussed in Fortner et al.\ 1989), one
being the disk, \rin~could be set by the accretion rate through the
disk (Kaaret 2000). In this picture, the inner disk radius moves in
when the accretion rate through the disk increases (e.g.  Miller et
al.\ 1998).  Eventually, if the NS radius is smaller than the radius
of the Innermost Stable Circular Orbit (ISCO at \Rms) predicted by
General relativity, the disk should stop at the ISCO and the kilo-Hz
QPO frequency should saturate.  Evidence for such a saturation at
\Rms~has been claimed for one source: 4U1820-303 (Zhang et al.\ 1998,
Bloser et al.\ 2000).  Although attractive, this scenario faces
already some problems.  First, it is unclear how systems with
luminosities differing by up to 2 orders of magnitude with QPO in the
same frequency range can keep the same accretion rate through the disk
at very different total accretion rate (Ford et al.\ 2000).
Furthermore, one would expect that at a constant kilo-Hz QPO frequency
(i.e. fixed disk mass accretion rate), the strength of the QPO should
decrease rapidly, when the X-ray count rate (tracing the total
accretion rate) increases.  Such a rapid decrease has not been
observed (M\'endez et al.\ 2001).  To conclude, timing and spectral
observations could be explained by a varying inner disk radius.
However, the fundamental parameter or the combination of parameters
that set the value of the inner disk radius has yet to be identified.

What is the nature of the hot inner accretion flow?  Driven by the
similarities with BHCs, despite the very different inner boundary
conditions (a solid surface as opposed to an event horizon) in the
hard spectral state, it has been hypothesized that the hot inner
accretion flow could be an ADAF (Barret et al.\  2000).  For the ADAF
to remain and for the hard spectra observed to be produced, the flow
must settle down on the NS surface in the form of hot optically thin
boundary layer (see computations in Narayan \& Yi 1995, Yi et al.\ 
1996).  At higher luminosities, in response to a larger cooling flux
from the disk (and possibly from a change in the nature of the
boundary layer), the ADAF should collapse.  Most of the emission could
then come from the boundary layer between the NS and the disk (e.g
Popham \& Sunyaev 2001, Inogamov \& Sunyaev 1999).  Although the
similarities between BHC and NS point to a similar accretion geometry
and emission processes, more theoretical work is needed to study the
nature of the inner flow and the structure and properties of such a
flow when it reaches the NS surface.

\subsection*{Non-thermal processes in NS}
In the above discussion, it is implicitly assumed that the main
emission mechanism is thermal Comptonization.  However, recent
observations have shown that NS, like BHC, may occasionally display
non-thermal state with hard X-ray spectra extending up to $\sim 200$
keV with no observable cutoffs (see Fig.  \ref{gx17+2}, for an early
review of non-thermal models specific to NSs, see Tavani \& Barret
1997).  For BHCs, these spectral states are generally interpreted in
the framework of two competing models.  The first one is involving
bulk motion Comptonization in a converging accretion flow (e.g.\ 
Titarchuk \& Zannias 1998).  This model is however unlikely to apply
to bright NS, because the radiation pressure caused by the large mass
accretion rate will prevent the flow from free-falling toward the NS
surface (Laurent \& Titarchuk 1999).  The second model is the hybrid
thermal/non thermal model, where a fraction of the accretion power
goes in the acceleration of non thermal electrons for the
Comptonization (Gierli{\'n}ski et al.\ 1999).  For Sco X-1, a jet is
directly and repeatedly observed with VLBA (e.g.  Bradshaw et al.\ 
1999) and the hard X-ray emission seems to correlate with periods of
radio flaring (Strickman \& Barret 2000).  In addition, two other
sources for which non thermal hard tails have been observed (Cyg X-2,
GX17+2) are already known to be all relatively bright radio sources,
likely generating radio-emitting outflows or jets (Fender \& Hendry
2000).  The fact that in general the radio emission is stronger on the
the so-called horizontal branch (e.g. Fender 2001), when the hard
X-ray tail was detected in GX17+2 (Di Salvo et al.\ 2000b) strongly
suggests that the hard X-ray and radio emissions are related.
Up-scattering of soft photons could therefore involve non-thermal
electrons from a jet (Di Salvo et al.\ 2000b).  Alternatively,
synchrotron emission from the jet itself could be responsible for the
variable hard X-ray tail observed (Markoff et al.\ 2000).  Further
simultaneous radio/hard X-ray observations are needed to test whether
the emission of a non thermal hard X-ray tail is indeed associated
with the formation of a jet both in low and high luminosity sources.
This would imply that jet formation can occur in systems covering a
broad range of accretion rates, and that the jet formation is more
likely related to changes of the accretion flow structure rather than
to a high accretion rate (Fender \& Hendry 2000).

\subsection*{Boundary layer models}
In the previous sections, the discussion draws on the similarities
between BH and NS accretion.  However, several models specific to NS
accretion have been proposed.  These models involve either a boundary
layer where a standard disk interacts with the NS surface (Inogamov \&
Sunyaev, 1999, and Popham \& Sunyaev, 2001) or an accretion gap
between a standard disk terminating at the ISCO and the NS surface
(e.g. Klu\'zniak \& Wagoner 1985).

If the accretion disk is not truncated and extends all the way to the
NS surface or to the last stable orbit, significant energy release is
expected to occur in the so-called boundary layer.  Sibigatullin \&
Sunyaev (2000) have recently computed approximated formulas for the
ratio of the boundary layer to the total luminosities for various NS
equations of states, and for various spin frequencies of the NS. For
plausible spin frequencies of the NS as inferred from kilo-Hz QPOs and
burst oscillations, it is found that the luminosity of the boundary
layer exceeds that of the disk by a large factor.  In this framework,
the hard component which dominates the source luminosity in NS could
arise from such a boundary layer, whereas the soft component would
originate from the accretion disk.  Inogamov \& Sunyaev (1999) have
modeled the boundary layer as an accretion belt around the NS equator.
The gas enters the spreading layer at nearly keplerian rotation
velocity.  The deceleration of the spreading matter occurs due to
friction against the dense underlying layer.  The energy release takes
place on the NS surface in a latitudinal belt whose width rises with
increasing luminosity.  Such a spread layer could be the seat of very
fast variability (1-2 kHz), and could therefore be responsible for the
extra high frequency noise component seen only in the power density
spectra of NS (Sunyaev \& Revnivtsev 2000). Alternatively, Popham \&
Sunyaev (2001) have modeled the boundary layer as part of the
accretion disk using the slim disk equations.  In this model, the drop
from the Keplerian to the stellar angular velocity takes places during
the radial inflow of the gas, rather than during the spread of matter
over the NS surface as in the above model. The energy is transported
by viscosity from the outer parts of the boundary layer to the more
slowly rotating inner parts, concentrating the energy release at the
NS surface. In both models, in the low \mdot~regime, relatively hard
spectra can be produced through thermal Comptonization of seed photons
emitted by the denser parts of the boundary layer. On the other hand,
at high \mdot~much softer spectra are predicted.  Finally, the last
model to be considered applies if the disk terminates at \Rms~and the
NS radius is smaller than \Rms~(as allowed for some soft NS equations
of state, Klu\'zniak \& Wagoner 1985). Beyond \Rms~the matter will
fall freely onto the surface of the NS.  The size of the gap will also
depend on the spin rate of the NS and the sense of rotation of the
disk versus the NS (Sibigatullin \& Sunyaev 2000).  The kinetic energy
of the fluid crossing the gap could be dissipated in the luminous
equatorial belt around the NS equator.  Self consistent computations
of the stucture of the belt showed that it could produce hard X-rays
for sufficiently low accretion rate, when the accretion gap is
optically thin to electron scattering (Klu\'zniak \& Wilson 1991, see
also Hanawa 1991).

Clearly the above boundary layer models make predictions that are in
qualitative agreement with the observations (e.g.\ hard X-ray emission
in the low \mdot~regime, softer spectra at higher \mdot).  Obviously,
it is unclear how this class of models could account for the
similarities between BH and NS accretion.  Furthermore, as we have
discussed above, in the hard spectral state, there is some evidence
that the accretion disk may be truncated at large radii, with the
nature and structure of the inner flow differing from that of a
standard disk.  How the properties of the inner flow will affect the
properties of the boundary layer remains to be explored.  On the other
hand, the above models might be more applicable in the high accretion
rate regime when the disk extends very close to the NS surface.

\section*{CONCLUSIONS}
To conclude, in NS like in the case of BHC, in the hard spectral
state, accretion could occur in the form of a truncated accretion disk
and a hot inner flow.  It is very tempting to relate the QPO frequency
and spectral changes to a varying inner disk radius, the hard spectral
state corresponding to the larger inner disk radius.  However the
parameter which makes the inner disk radius vary has yet to be
identified. Similarly the nature of the hot inner flow is unknown. By
analogy with BHCs, it could be an advection dominated accretion flow,
settling on the NS surface in the form of a hot optically thin
boundary layer.  At higher accretion rate, the inner accretion flow
would cool down and collapse, and as the disk would get closer to the
NS, most of the emission could arise from a classical boundary layer.
Like in BHCs, non thermal processes occur also in NS. The evidence is
growing that mass outflows are also important in NS, and that the non
thermal hard X-ray emission is related to periods of radio emission,
and hence could be associated with the formation of a jet.

Broad band observations have been shown to be a powerful tool to study
the accretion flows around compact objects.  Similarly, when these
observations are correlated with fast timing observations, very strong
constraints can be derived. Significant progresses have been possible
recently thanks to \sax~and \rxte.  Soon, more progresses should be
accomplished with \integral~which will have both broad band coverage
and higher sensitivity in hard X-rays, and later with a next
generation X-ray timing satellite which should provide an order of
magnitude improvement in sensitivity for timing studies together with
enhanced broad band spectral capabilities (e.g.\ the proposed
Experiment for X-ray Timing and Relativistic Astrophysics, Barret et
al.\ 2001).
\section*{ACKNOWLEDGEMENTS}
It is my pleasure to thank Dr.  B. Czerny for her support during the
preparation of this paper.  I am grateful to T. Di Salvo for providing
me with the unfolded spectrum of GX17+2, to C. Done, R. Fender, M.
Gierli{\'n}ski, M. M\'endez, L. Natalucci, J.F.  Olive, T.
Oosterbroek, T. Di Salvo for helpful comments along the preparation of
this paper.

\end{document}